\documentclass[preprint,showpacs,aps,prd,nofootinbib,floatfix,amsmath,amssymb]{revtex4}
\usepackage{graphics}
\usepackage{amsmath}
\usepackage{color}
\usepackage{subfig}
\usepackage{graphicx}

\begin{document}
\title{Chaos in the Y-chromosome evolution?}

\author{Matheus P. Lobo }%
\email{mplobo@uft.edu.br}
\affiliation{
Universidade Federal do Tocantins - Campus Universit\'ario
de Aragua\'\i na\\
Av. Paraguai (esquina com Urixamas), Cimba \\ Aragua\'\i na - TO, 77824-838, Brazil
}

\author{Edison T. Franco}%
\email{edisonfranco@uft.edu.br}
\affiliation{
Universidade Federal do Tocantins - Campus Universit\'ario
de Aragua\'\i na\\
Av. Paraguai (esquina com Urixamas), Cimba \\ Aragua\'\i na - TO, 77824-838, Brazil
}

\author{Nilo M. Sotomayor}%
\email{nmsch@uft.edu.br}
\affiliation{
Universidade Federal do Tocantins - Campus Universit\'ario
de Aragua\'\i na\\
Av. Paraguai (esquina com Urixamas), Cimba \\ Aragua\'\i na - TO, 77824-838, Brazil
}

\author{Felipe R. Costa}%
\email{offelipecosta@gmail.com}
\affiliation{
Universidade Federal do Tocantins - Campus Universit\'ario
de Aragua\'\i na\\
Av. Paraguai (esquina com Urixamas), Cimba \\ Aragua\'\i na - TO, 77824-838, Brazil
}

\date{\today}
\begin{abstract}
The Y-chromosome degeneration is still an intriguing mechanism and comprises the very origin of sex. We present a coupled version of the well known logistic map and the logistic equation describing the evolution of XY chromosomes. Although chaos was found in X, Y chromosomes do not evolve chaotically. A mathematical constraint is shown as the responsible for this behaviour. In addition, analytical solutions are presented for the differential equations herein.
\end{abstract}


\maketitle
\section{Introduction}

The Y chromosomes are genetically degenerated, smaller in size, with less functional genes than their X-chromosomes partners \cite{charles}.

Several models were suitably presented as mechanisms to serve as a reinforcement \cite{lb1} or to even originate Y degeneration itself \cite{lb2}. Additionally, the predominance of X chromosomes is also responsible for the shrinking of the Y chromosomes during millions of years of evolution \cite{stf1}.

There are many forces acting simultaneously on the sex-chromosomes evolution. LOBO and ONODY showed that the ``no-recombination" argument (for XX chromosomes) is not necessary and still leads to Y-chromosome degeneration \cite{lb1}. BIECEK and CEBRAT showed that Y chromosomes shrink due to a weaker selection pressure acting on Y \cite{cb1}.

Here we present a modified logistic map for XY-chromo\-somes evolution. The ecologist Robert May showed that the logistic equation is a simple model with rich dynamics \cite{rmay1,rmay2} and it has been considered by the mathematician Ian Stewart one of the 17 equations that changed the world \cite{ian}. Despite its simplicity, it can generate deterministic chaos \cite{groff} and its generalization can describe a number of different biological species \cite{ex1,ex2}.

The present paper is organized as follows.
In Sec.~\ref{sec:model} we study the structure of the coupled logistic map for XY chromosomes. In Sec.~\ref{sec:butterfly} we show how the initial conditions affects the number of chromosomes and the number of individuals. In Sec.~\ref{sec:constraints} we analyse the mathematical constraints which prevents the chaotic behavior of Y chromosomes. In Sec.~\ref{sec:analytical} we set the analytical description and the fundamental ideas of our model for XY-chromosomes evolution. Our conclusions and the discussion of our results are treated in Sec.~\ref{sec:discussion}.

\section{XY-Logistic Model}
\label{sec:model}

We modified the logistic map in order to describe the mathematical asymmetry of the XY chromosomes and their coupled properties as well. The discrete equations read as
\begin{eqnarray}
x_{i+1}&=&rx_i\left(1-\frac{x_i}{3k}\right),\label{mx}\\
y_{i+1}&=&r\frac{\left(\frac{x_i}{3}+y_i\right)}{2}\left[1-\frac{\left(\frac{x_i}{3}+y_i\right)}{2k}\right],\label{my}
\end{eqnarray}

\noindent where $r$ is the fertility, $k$ is the population threshold, $x_i$ and $y_i$ stands for the number of X and Y chromosomes, respectively. Equation (\ref{mx}) is the traditional logistic map, here applied for the X chromosomes. Equation (\ref{my}) is the modified logistic map for the Y chromosomes. These are non-overlapping populations. The term $\frac{x_i}{3}$ in equation (\ref{my}) accounts for the proportion of sexual chromosomes, 3X for each Y chromosome. Note that the maximum number of X (Y) chromosomes is limited to $3k$ ($k$). Our equations implicitly assume that the same number of males and females will be reached in the steady state. We assume $k=1000$ hereafter.

Our model for the population growth is based solely in the XY chromosomes. In the steady state we expect that the number of males (XY) and females (XX) should be approximately the same. In terms of the number of chromosomes, it means that X will be three times greater than Y chromosomes.

The corresponding number of males ($N^m$) is given by the number of XY pairs of chromosomes. For $x>y$, the number of males is equal to $y$, since we assume that all of the Y chromosomes can be combined with their X partners. On the other hand, if $y>x$, the maximum number of males is limited to $x$, as summarized by
\begin{eqnarray}\label{eq:NmDif}
N^m=
\begin{cases}
  y & \text{for } x>y \\
  x & \text{for } x<y,
\end{cases}
\end{eqnarray}

\noindent and the number of females ($N^f$) will be given by the half of the remaining X chromosomes, as follows
\begin{eqnarray} \label{eq:NfDif}
N^f=
\begin{cases}
  \frac{1}{2}(x-y) & \text{for } x>y \\
  0 & \text{for } x<y.
\end{cases}
\end{eqnarray}

\begin{figure}
\center
\subfloat[\label{fig:Aa}]{\includegraphics[height=0.30\textwidth]{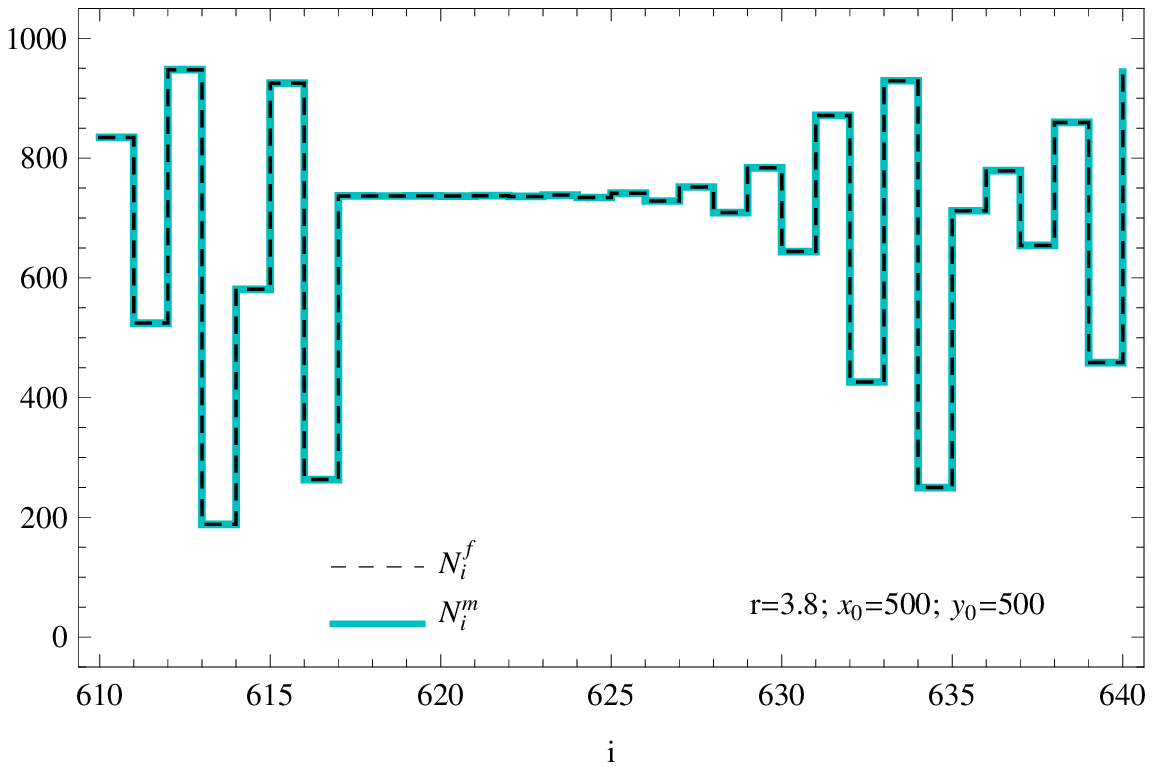}}
\hfill
\subfloat[\label{fig:Ab}]{\includegraphics[height=0.30\textwidth]{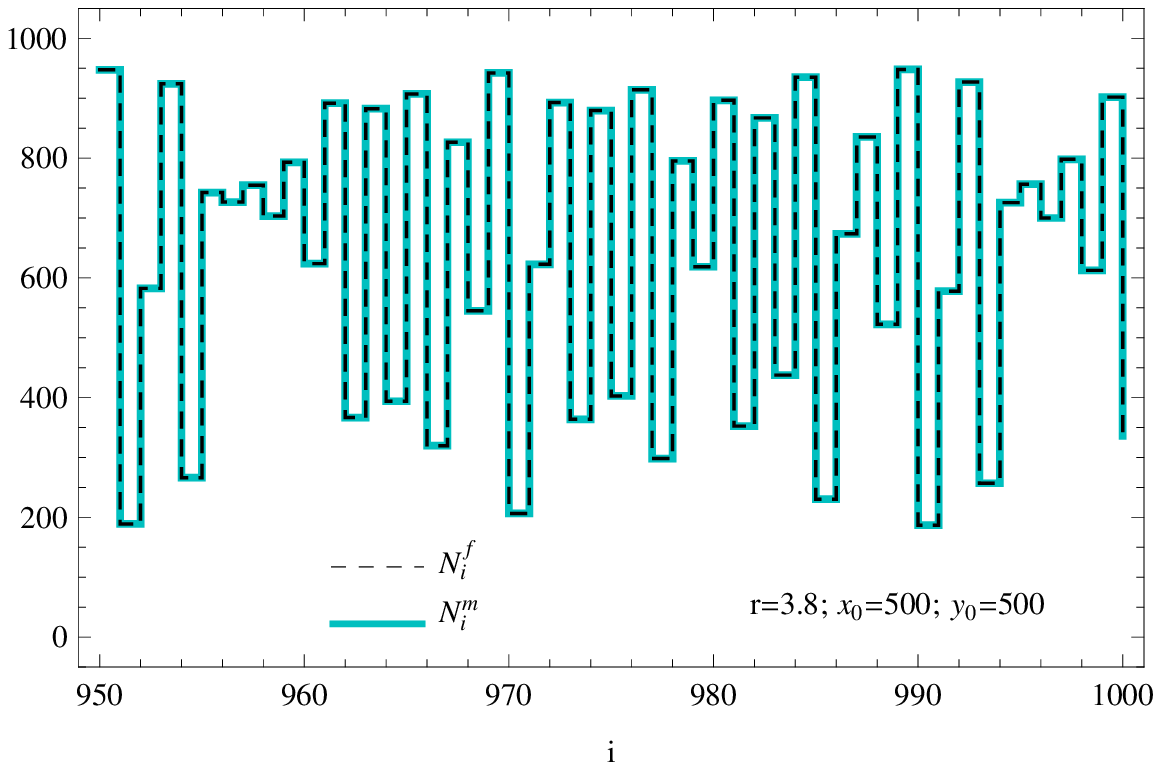}}
\caption{Evolution of the number of males ($N^m$) and females ($N^f$) after (a) 610 and (b) 950 generations. The fixed parameters are: $r=3.8$, $x_0=500$, and $y_0=500$.}
\label{fig:A}
\end{figure}

Equations (\ref{eq:NmDif}) and (\ref{eq:NfDif}) provides the maximum number of males in a given population of X and Y chromosomes. Figure \ref{fig:A} shows that the population of males and females are equal and stable. Although the proportion of males and females are roughly the same, X chromosomes appears 3 times more often than Y, as predicted by equations (\ref{mx}) and (\ref{my}).

Surprisingly, Figure \ref{fig:A}(a) shows that nearly after 618 generations there is an equal number of males and females for about 10 generations. We have seen that this effect happens only once until the thousandth generation. We are exploring this effect in depth and it should be published elsewhere.

\section{Chaos and the butterfly effect}
\label{sec:butterfly}

\begin{figure}
\center
\subfloat[\label{fig:Ba}]{\includegraphics[height=0.30\textwidth]{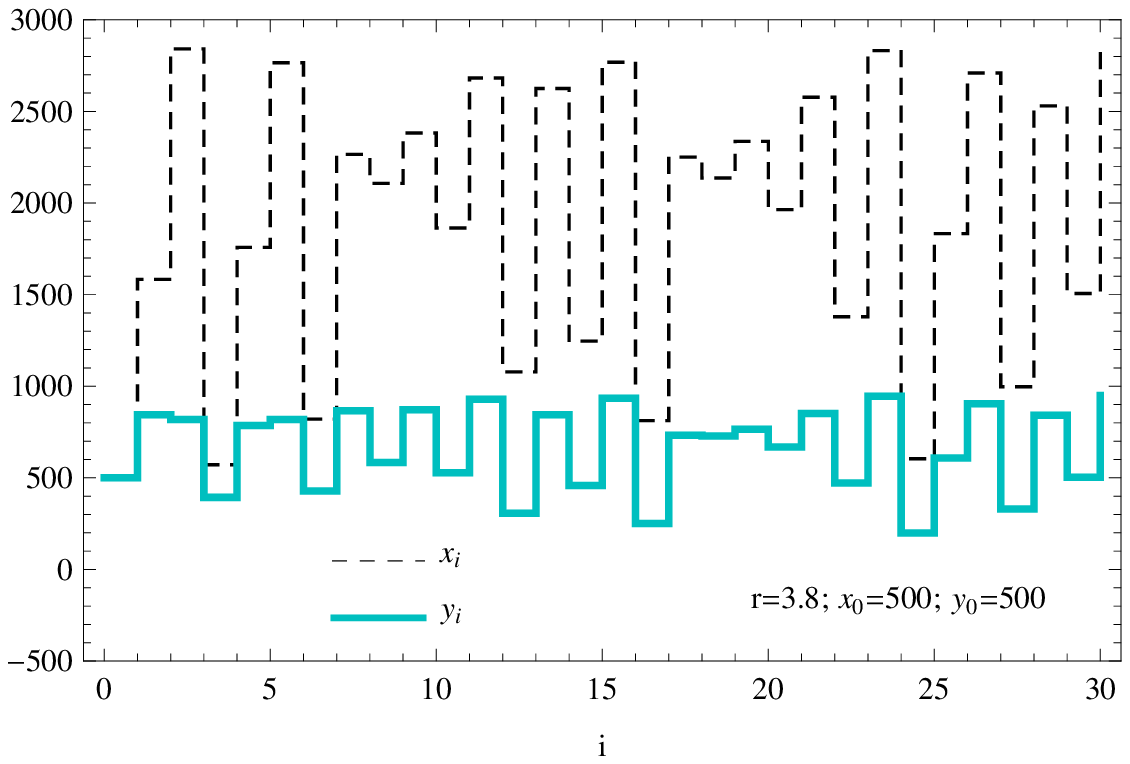}}
\hfill
\subfloat[\label{fig:Bb}]{\includegraphics[height=0.30\textwidth]{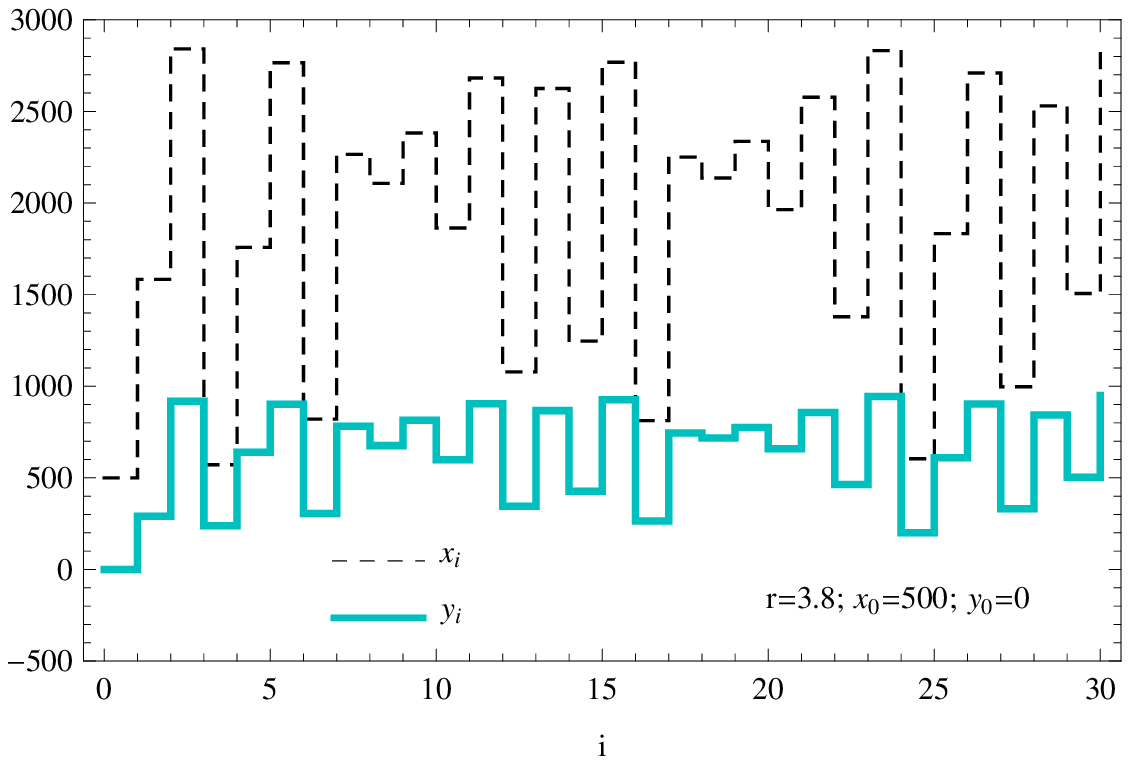}}
\hfill
\subfloat[\label{fig:Bc}]{\includegraphics[height=0.30\textwidth]{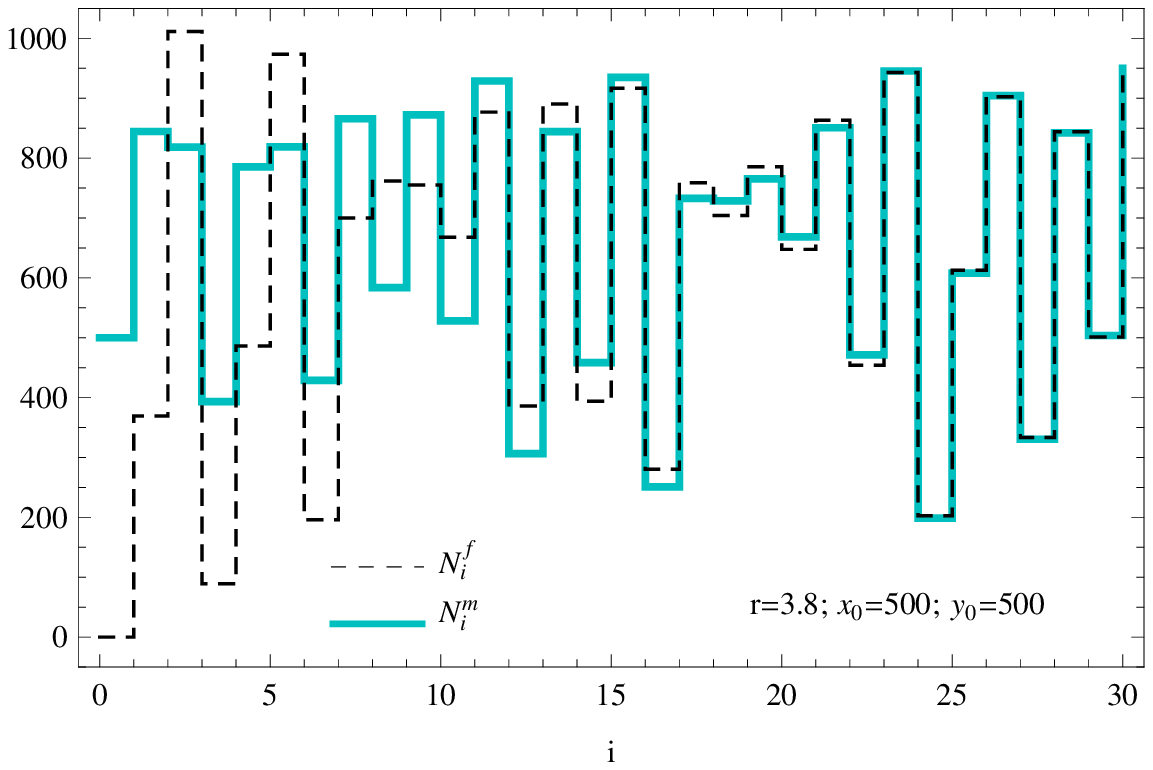}}
\hfill
\subfloat[\label{fig:Bd}]{\includegraphics[height=0.30\textwidth]{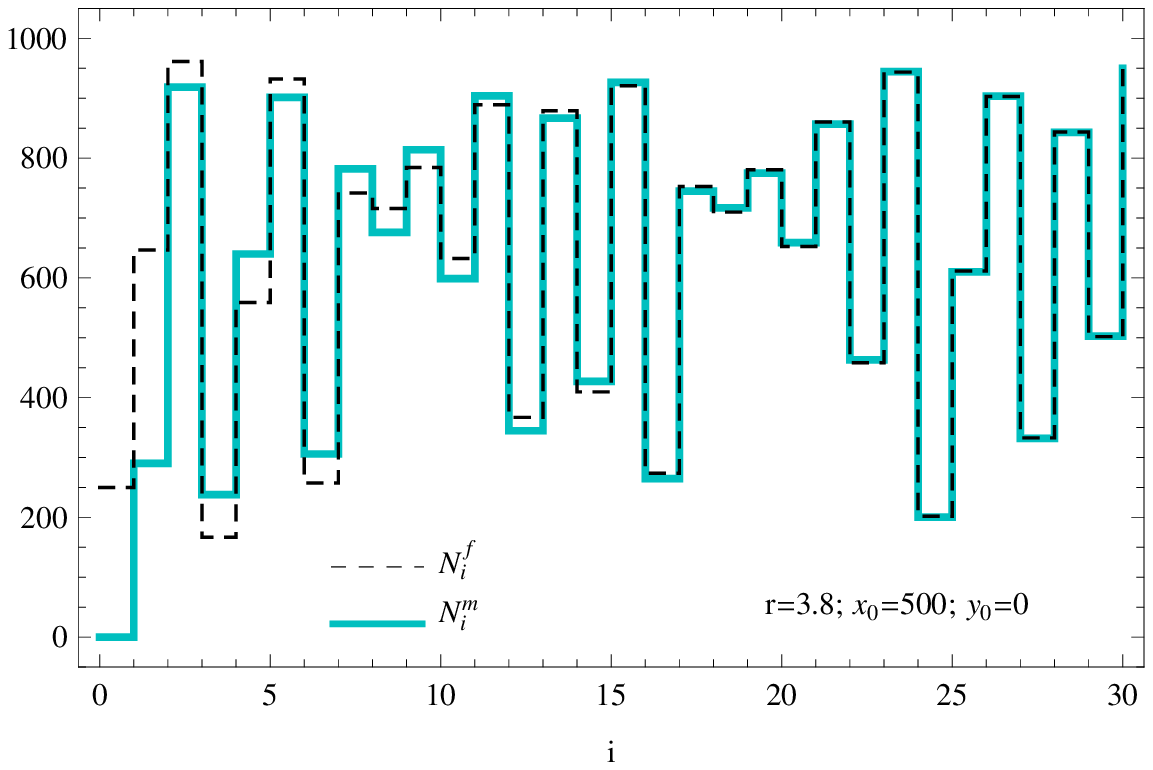}}
\caption{Evolution of $x_i$ and $y_i$ for (a) $y_0=500$ and (b) $y_0=0$. Trajectories of $N^m$ and $N^f$ for (c) $y_0=500$ and (d) $y_0=0$. The fixed parameters are: $r=3.8$, and $x_0=500$.}
\label{fig:B}
\end{figure}

In this section we have iterated the maps (\ref{mx}) and (\ref{my}) in order to show the sensibility of the system to the initial conditions regarding the population size. We also show the Lyapunov expoents for X and Y chromosomes. To pinpoint the differences on the population size, Figure \ref{fig:B} shows the trajectories of males, females, X and Y chromosomes over the generations ($i$).

We varied the initial condition on $y_0$ with the other parameters fixed. Figure \ref{fig:B}(a) and \ref{fig:B}(b) shows the evolution of the X and Y chromosomes, and Figure \ref{fig:B}(c) and \ref{fig:B}(d) shows the trajectories of the number of males and females. In all the four cases we see that the initial condition on Y had little effect only on the first 10 (a and b) and 20 generations (c and d). This is the first sign that the system is not sensible on the initial conditions of Y.

\begin{figure}
\center
\subfloat[\label{fig:Ca}]{\includegraphics[height=0.30\textwidth]{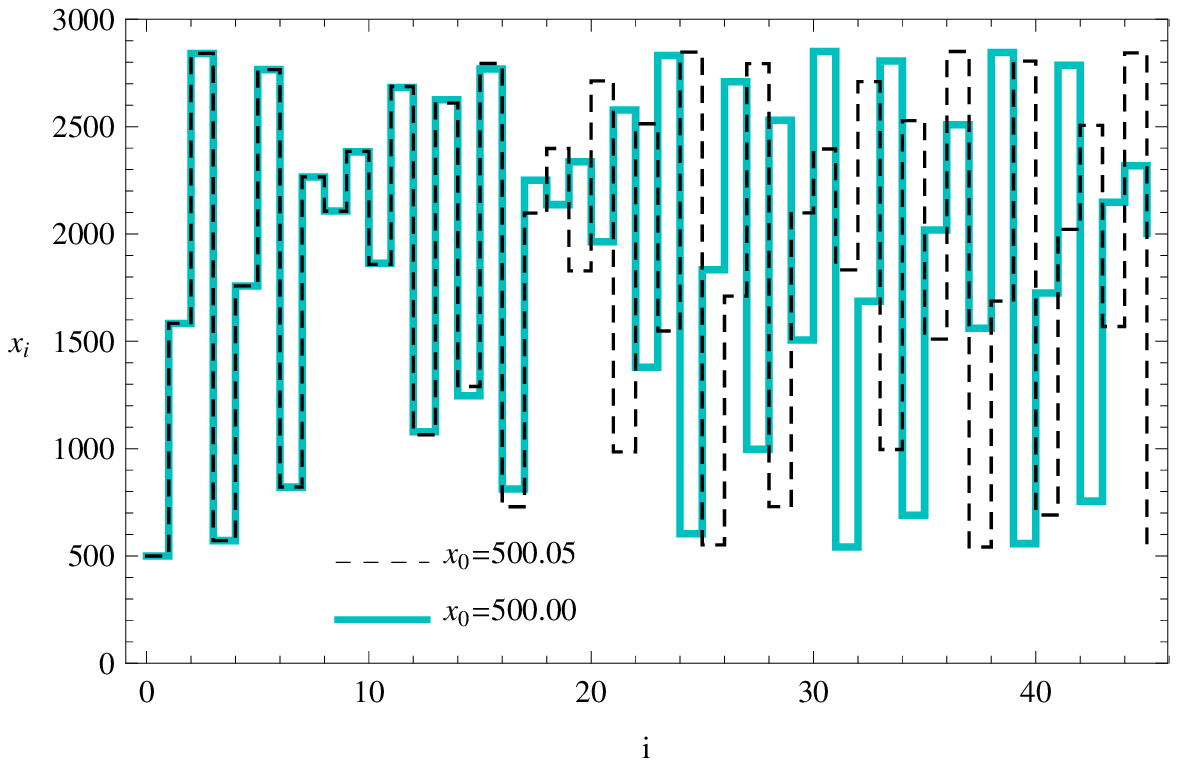}}
\hfill
\subfloat[\label{fig:Cb}]{\includegraphics[height=0.30\textwidth]{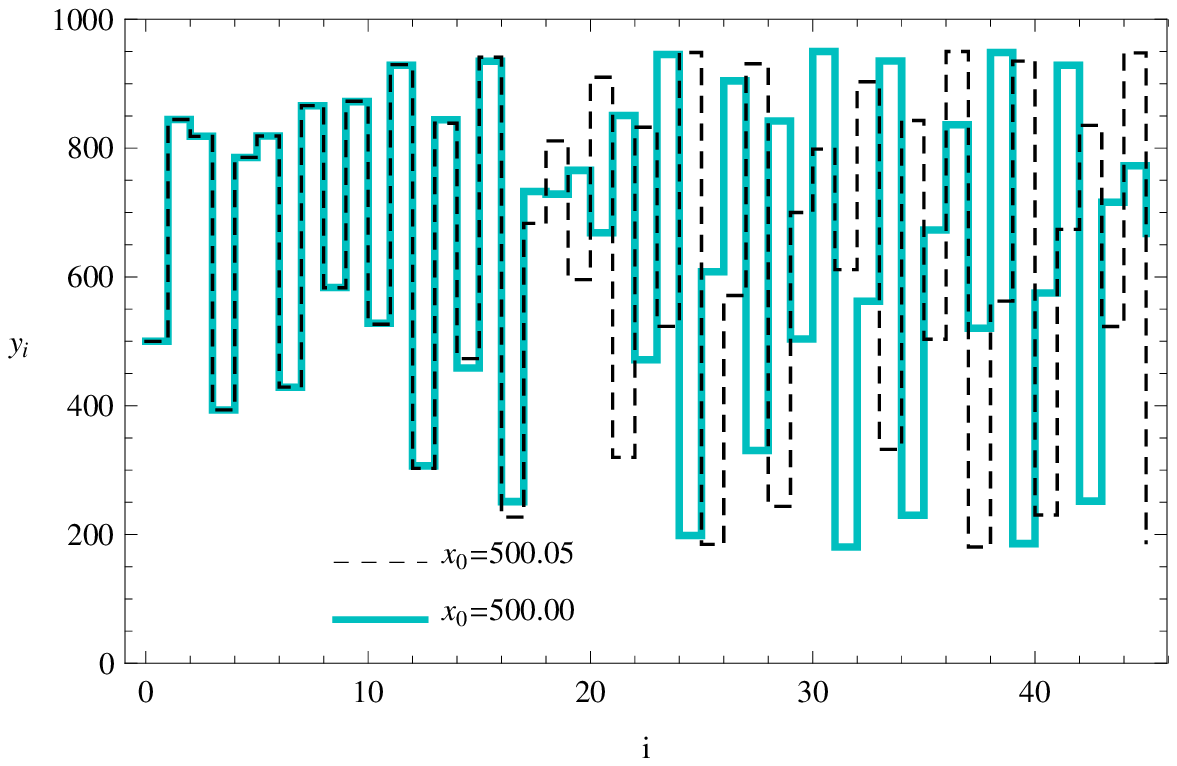}}
\hfill
\subfloat[\label{fig:Cc}]{\includegraphics[height=0.30\textwidth]{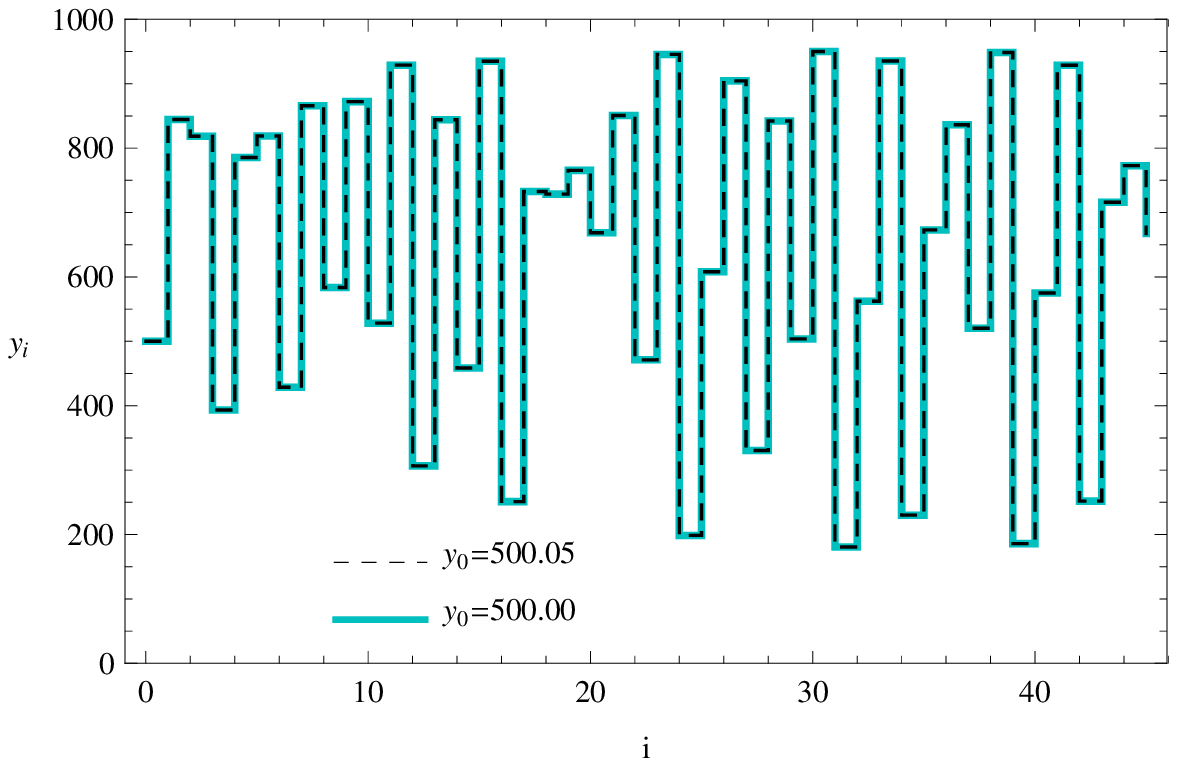}}
\caption{Trajectories for the XY-Logistic map [Eqs. (\ref{mx}) and (\ref{my})] with $x_0=500$ (solid) and $x_0=500.05$ (dashed) showing the evolution of (a) X, and (b) Y chromosomes; and with $y_0=500$ (solid) and $y_0=500.05$ (dashed) for the Y-chromosomes evolution.}
\label{fig:C}
\end{figure}

Chaos is the result of a very rich dynamics. The damped oscillations and the void of steady state are necessary (although not sufficient) conditions for chaotic regimes \cite{hastings}. The butterfly effect can be seen here (on X chromosomes) by means of the dynamics of the population growth regarding different initial conditions. We varied the initial conditions in order to verify the butterfly effect. Figure \ref{fig:C} shows the sensitivity of the XY-logistic model in the initial population size of X chromosomes. It is no surprise, however, that X exhibits chaos, since it represents the well-known logistic map.

\begin{figure}[!h]
\begin{center}
    \includegraphics[height=0.30\textwidth]{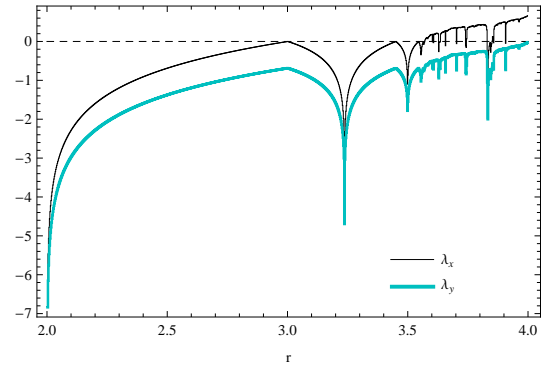}
\caption{Lyapunov exponents for the XY-Logistic Model found for $r$ between 2 and 4.}
\label{fig:D}
\end{center}
\end{figure}

The ultimate hallmark of chaos is shown by the Lyapunov exponents for it is the quantity that characterizes the rate of separation of infinitesimally close trajectories \cite{strogatz}. Figure \ref{fig:D} shows the Lyapunov exponents (LE) for both X and Y chromosomes. A positive LE is an indication of chaotic regimes. Surprinsingly, we have $\lambda_x>0$ for a portion of the fertility spectrum and $\lambda_y<0$ for the whole fertility range studied. This indicates that the Y chromosomes escape from the chaotic regime.

\section{Mathematical constraints}
\label{sec:constraints}
The exhibition of chaos in the logistic map is intrinsically related to the shape of the corresponding parabola,
\begin{equation}
x_{i+1}=-\frac{r}{3}x_i^2+rx_{i},
\label{parx}
\end{equation}

\noindent where the values of $r$ determines the chaotic region \cite{groff}. We considered $k=1$ for the sake of simplicity. Equation (\ref{my}) can also be written as
\begin{equation}
y_{i+1}=-ay_i^2+b_iy_i+c_i,
\label{pary}
\end{equation}

\noindent where $a=\frac{r}{4}$, $b_i=\frac{r}{2}-\frac{r}{6}x_i$ and $c_i=\frac{r}{6}x_i-\frac{r}{36}x_i^2$. Whether there is chaos in the Y chromosome's evolution can be checked comparing equations (\ref{parx}) and (\ref{pary}), i.e., the parabola for Y is the same parabola for X if

\[a=\frac{r}{3}, \hspace{0,6cm} b_i=r, \hspace{0,6cm} c_i=0.\]

\noindent Since $a=\frac{r}{4}$, the parabola for Y cannot be chaotic. In addition, the condition $b_i=r$ leads to the nonphysical result, $x_i<0$. Negative numbers are not allowed in the discrete spectrum of any population model and of course in the XY-Logistic Model. Therefore we conclude that the parabola for the Y chromosomes never behaves within the chaotic regime such as in the X chromosomes.

\section{Analytical XY-Chromosomes Evolution}
\label{sec:analytical}

The classical Verhulst's model for the population change can be described by the differential equation
\begin{equation}
\frac{dN}{dt}=rN\left( 1-\frac{N}{k}\right),
\end{equation}%

\noindent where $N$ is the number of individuals in the population, $r$ is the Malthusian growth rate and $k$ is the carrying capacity of the environment of a certain type of individuals.

The differential versions of the equations (\ref{mx}) and (\ref{my}) are given by
\begin{eqnarray}
\frac{dx}{dt} &=&rx\left(1-\frac{x}{3k}\right),  \label{eq:DifX} \\
\frac{dy}{dt} &=&\frac{r}{2}\left(\frac{x}{3}+y\right) \left[ 1-\frac{1}{2k}\left(\frac{x}{3}+y\right) \right].  \label{eq:DifY}
\end{eqnarray}

\begin{figure}
\center
\subfloat[\label{fig:Ea}]{\includegraphics[height=0.30\textwidth]{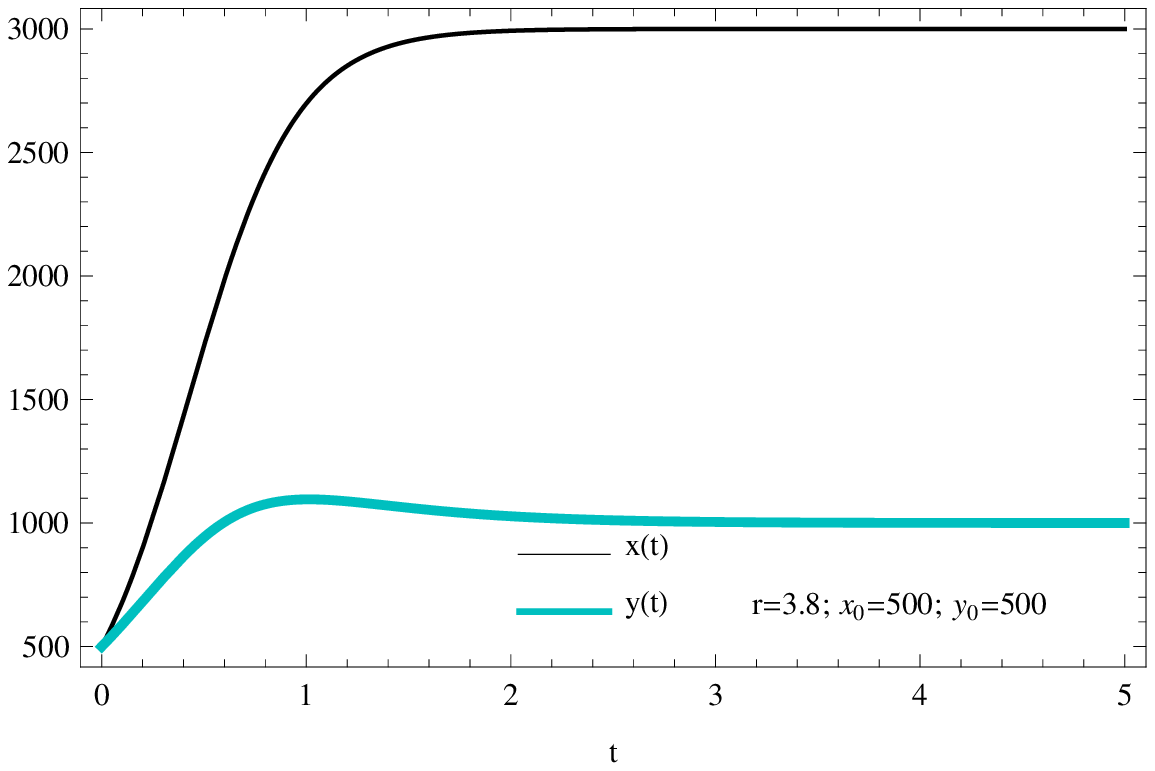}}
\hfill
\subfloat[\label{fig:Eb}]{\includegraphics[height=0.30\textwidth]{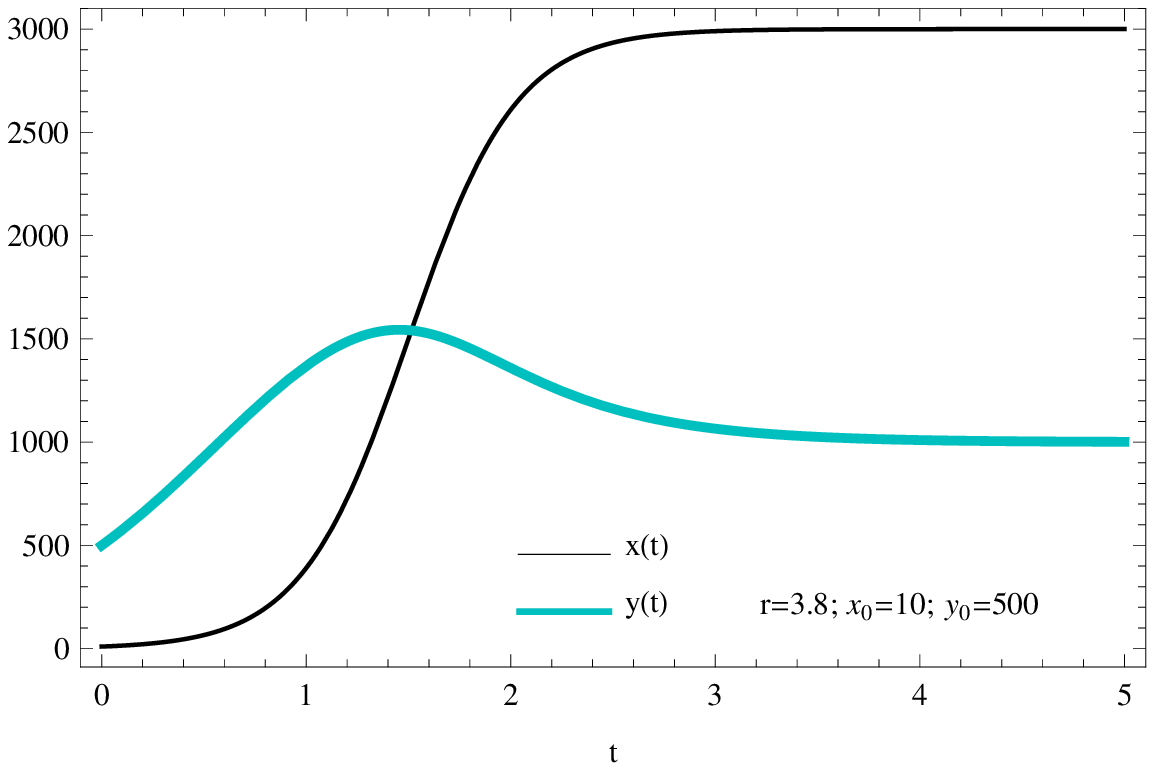}}
\caption{Analytical evolution of the X and Y chromosomes using Eqs. (\ref{eq:xt1}) and (\ref{eq:yt1}).}
\label{fig:E}
\end{figure}

The solution of Eqs. (\ref{eq:DifX}) and (\ref{eq:DifY}) are
\begin{equation}
x\left( t\right) =\frac{3k\mathrm{e}^{rt}}{\xi ^{2}+\mathrm{e}^{rt}}, \label{eq:xt1}
\end{equation}%
and
\begin{eqnarray}
y\left( t\right) &=&-\frac{2k\lambda \mathrm{e}^{\frac{rt}{2}}}{\left( \xi ^{2}+%
\mathrm{e}^{rt}\right) \left\{ \lambda \left[ \cot^{-1}\xi -\cot^{-1}%
\left( \xi \mathrm{e}^{-\frac{rt}{2}}\right) \right] +2\xi \right\} }\nonumber\\
&+&\frac{k\mathrm{e}^{rt}}{\left( \xi ^{2}+\mathrm{e}%
^{rt}\right) }\label{eq:yt1},
\end{eqnarray}

\noindent where $\xi \equiv \sqrt{\frac{3k-x_{0}}{x_{0}}}\ $\ and $\lambda \equiv
\left( \frac{x_{0}-3y_{0}}{x_{0}}\right) $. In Figure \ref{fig:E} we have shown chromosomes evolution with different initial conditions.

Notice that for greater values of $x_0$, the faster is the stabilization of $x$ and $y$, i.e., there are necessary less interactions to reach saturation. Conversely, with small values of $x_0$, the number of $y$ oscillates and have a higher maximum than with greater $x_0$. There is a special initial condition, $x_{0}=3y_{0}\ $, in which $\lambda =0$ and thus
\begin{equation}
x\left( t\right) =\frac{3kx_{0}\mathrm{e}^{rt}}{3k+x_{0}\left( \mathrm{e}%
^{rt}-1\right) }=3y\left( t\right).   \label{eq:xtyt}
\end{equation}%
\noindent which keeps the homogeneous relation $x=3y$ for any time $t$.

In this vein, it is interesting to observe how is the evolution of $y$ in terms of $x$ to check the proportion of the number of males and females. Considering the implicit function $y\left( x\left( t\right) \right)$ we have from Eq. (\ref{eq:DifY}),
\begin{equation}
\frac{dy}{dx}\frac{dx}{dt}=\frac{r}{2}\left( \frac{x}{3}+y\right) \left[ 1-%
\frac{1}{2k}\left( \frac{x}{3}+y\right) \right].
\end{equation}

\begin{figure}
\center
\subfloat[\label{fig:Fa}]{\includegraphics[height=0.30\textwidth]{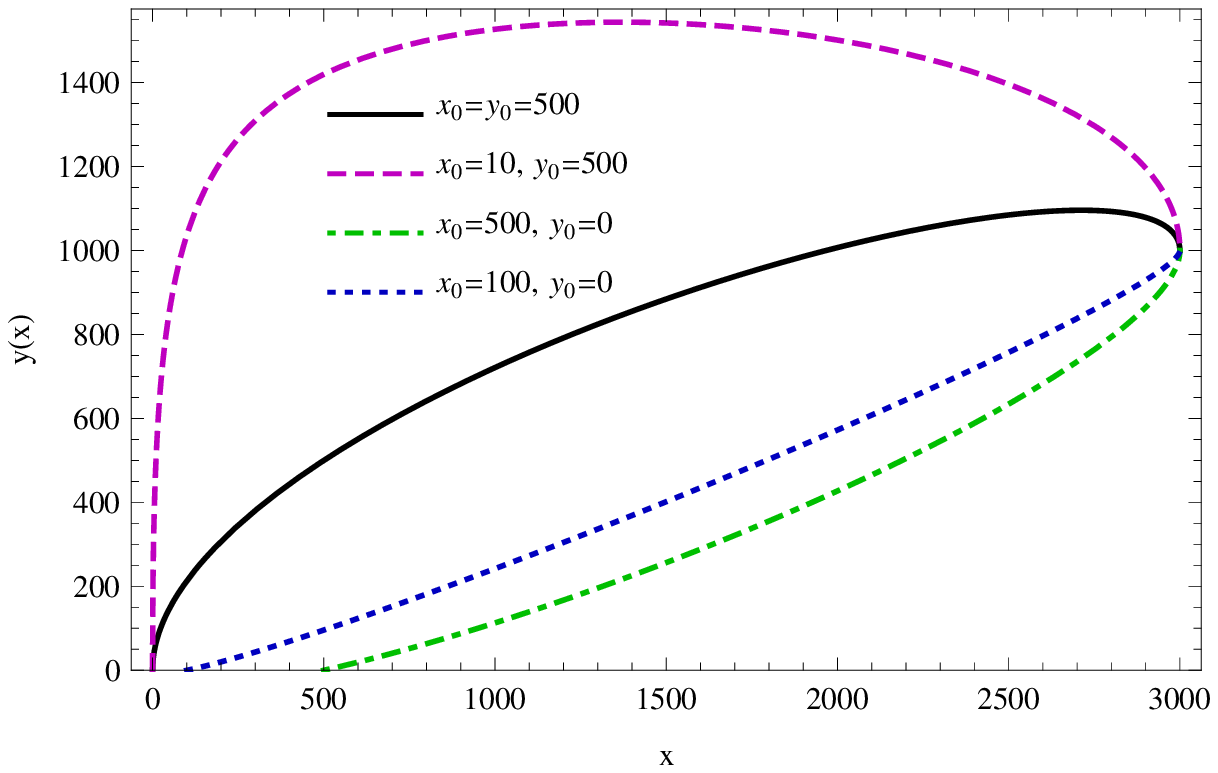}}
\hfill
\subfloat[\label{fig:Fb}]{\includegraphics[height=0.30\textwidth]{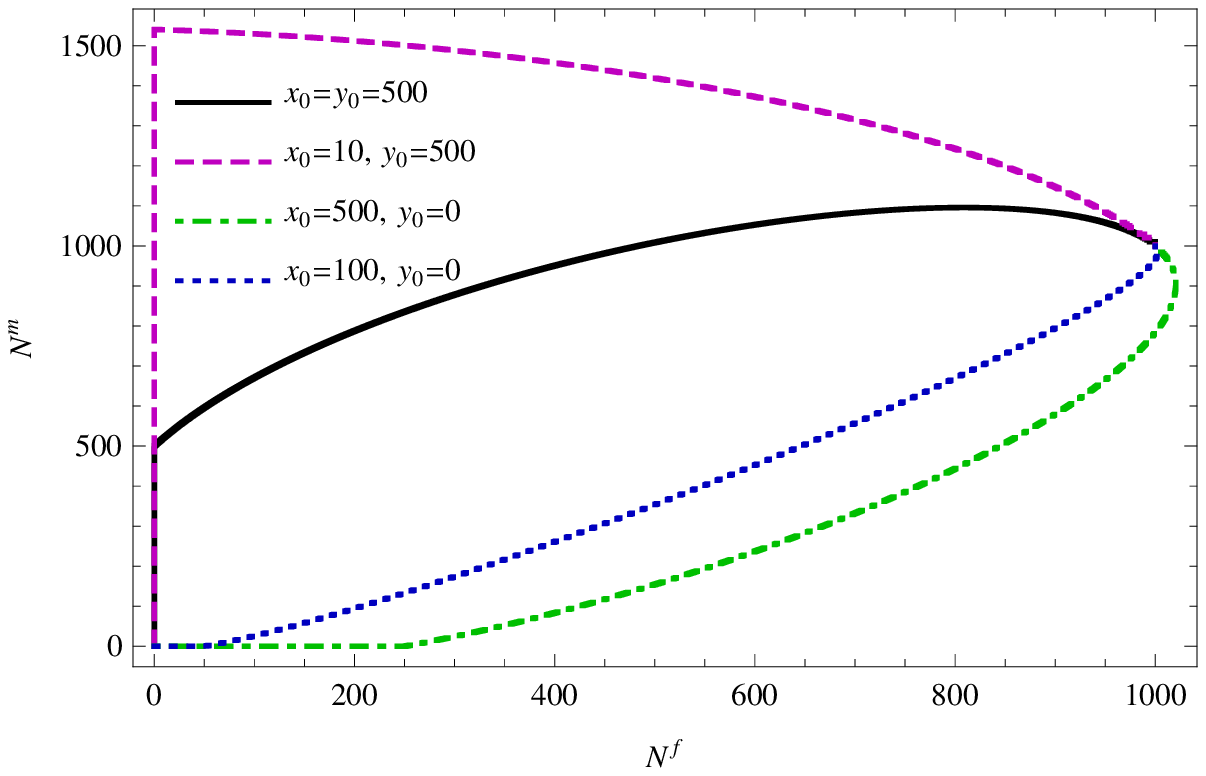}}
\caption{Analytical evolution of (a) $y(x)$  using Eq. (\ref{eq:yxalg}) and of (b) the number of males in terms of females using Eqs. (\ref{eq:NmDif}) and (\ref{eq:NfDif}).}
\label{fig:F}
\end{figure}

\noindent Using Eq. (\ref{eq:DifX}) we find the analytical solution for $y\left( x\right)$ as
\begin{equation}
y\left( x\right) =\frac{x}{3}+\frac{\rho _{0}\sqrt{\left( 3k-x\right) x}}{1-\frac{3}{2}\rho _{0}\left(\tan ^{-1}\sqrt{\frac{3k}{x}-1}-\tan ^{-1}\sqrt{\frac{3k}{x_{0}}-1}\right) },  \label{eq:yxalg}
\end{equation}

\noindent where $\rho _{0}\equiv \left( y_{0}-\frac{x_{0}}{3}\right) \frac{1}{\sqrt{\left( 3k-x_{0}\right) x_{0}}}$. The complicated relation between $y$ and $x$ given by Eq. (\ref{eq:yxalg}) is exemplified by the numerical solutions in the Figure \ref{fig:F}, where we have shown the behavior of the X and Y chromosomes evolution for different initial conditions. Figure \ref{fig:F}(a) shows that for small $x_0$, $y$ grows very fast. This means that the universe is dominated by Y until some X chromosomes are free to combine and produce females. As $x$ grows, the saturation is reached at $(x,y)=(3\,000,1\,000)$.

\begin{figure}
\center
\subfloat[\label{fig:Ha}]{\includegraphics[height=0.30\textwidth]{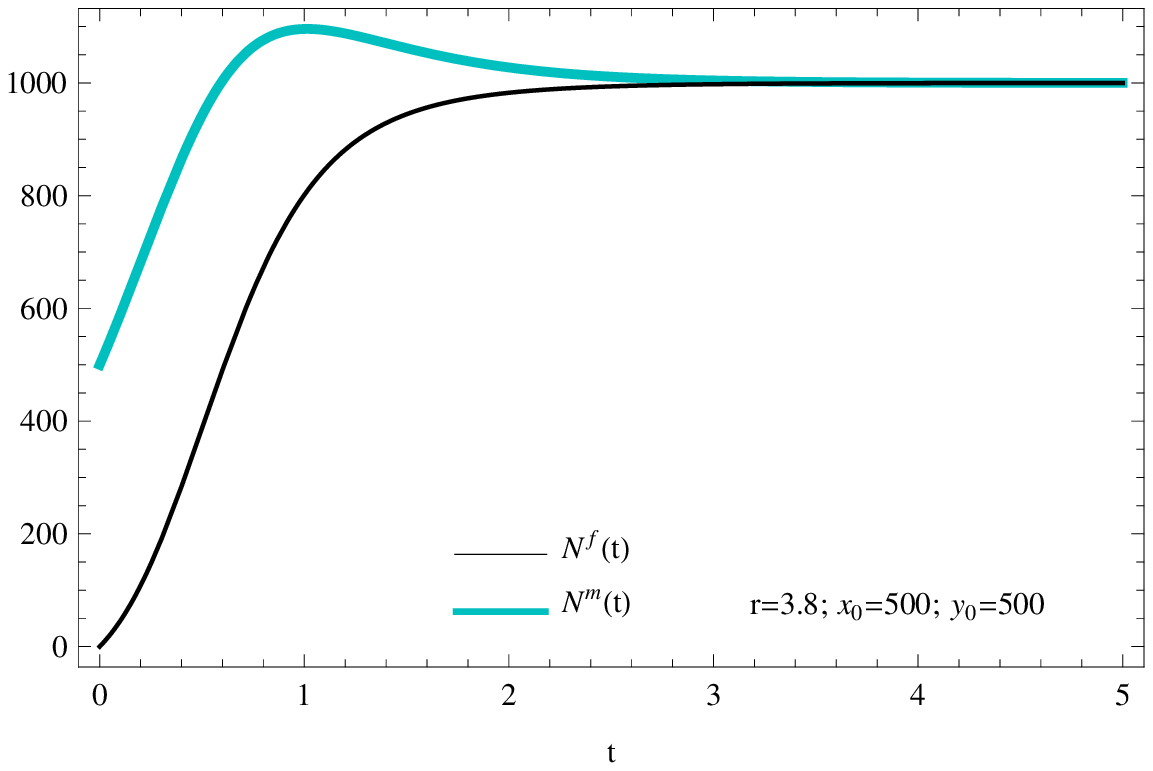}}
\hfill
\subfloat[\label{fig:Hb}]{\includegraphics[height=0.30\textwidth]{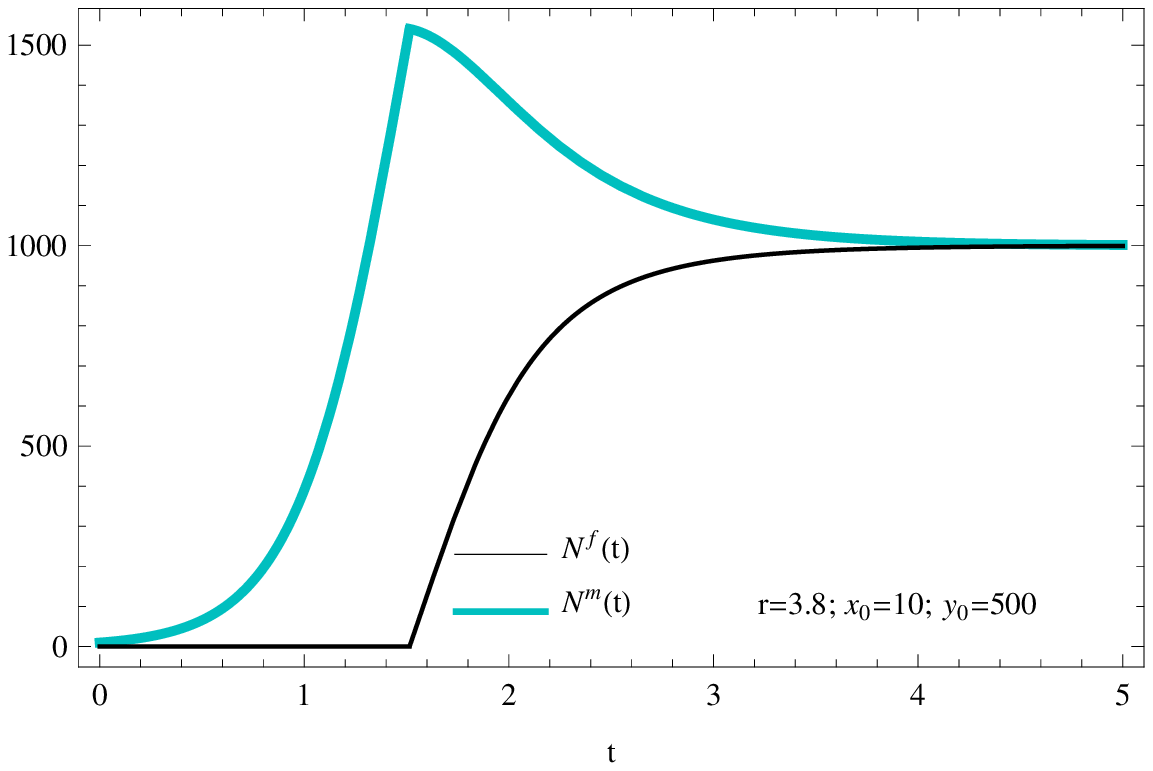}}
\caption{Analytical evolution of X and Y chromosomes using Eqs. (\ref{eq:xt1}) and (\ref{eq:yt1}).}
\label{fig:H}
\end{figure}

The Figure \ref{fig:H} shows the time evolution of chromosomes and the (maximum) number of males and females. The initial state, $x_0=y_0=500$, allows from the very beginning ($t=1$) the population with the maximum number of males (Figure \ref{fig:H}(a)): for each Y chromosome there is an X chromosome and thus the total number of males matches the number of Y chromosomes. On the other hand, the initial condition $x_0=10,\,y_0=500$ (Figure \ref{fig:H}(b)) populates the environment with many free Y chromosomes, which are not combined due to the small initial $x=x_0$ chromosomes. Thus, in this case, until $y$ is greater than $x$, the number of males is equal to $x$ and still growing without the creation of any female individuals. When $x$ becomes greater than $y$, nearly at $t\approx 1.6$, the saturation of male individuals is obtained and it starts to decrease while the number of females starts to grow. Nearly at $t\approx 4.0$ both number of males and females stabilizes and are approximately the same since $x$ and $y$ reach the saturation limit at $3\,000$ and $1\,000$, respectively.

\section{Discussion}
\label{sec:discussion}

We have showed that chaos is not present on Y-chromosomes evolution when it is coupled with X chromosomes. An heuristic explanation for the Y-chromosome degeneration by means of deterministic chaos is given as follows. The original Y appeared in some mammals (therian) approximately 180 million years ago \cite{cortez}. Small differences in initial conditions occured on the primeval X chromosomes -- due to quantum fluctuations in the light of quantum biology \cite{philip,lambert} -- might have led to the sex-determining genes.

We showed that another mathematical asymmetry, in addition to \cite{lb2}, might in turn be responsible for the Y-chromosome degenarion and to the origin of the sex-determ\-ining mechanisms. The butterfly effect has printed a permanent signature in the course of their evolution in the primeval era.

%

\end{document}